\documentclass[aip,pof,preprint,groupedaddress]{revtex4-1}
\usepackage{graphicx}
\usepackage{amssymb}

\begin{document}

\title{Temperature fluctuations in a changing climate: an ensemble-based experimental approach}

\author{Miklos Vincze$^{1,2}$}
\author{Ion Borcia$^{3}$}
\author{Uwe Harlander$^{3}$}

\affiliation{$^1$von K\'arm\'an Laboratory of Environmental Flows;}
\affiliation{$^2$MTA-ELTE Theoretical Physics Research Group}
\affiliation{$^3$Department of Aerodynamics and Fluid Mechanics, Brandenburg University of Technology, Cottbus-Senftenberg}


\begin{abstract}
There is an ongoing debate in the literature about whether 
the present global warming is increasing local and global temperature variability \cite{ref1}.
The central methodological issues of this debate relate to the proper treatment of normalised temperature anomalies 
and trends in the studied time series which may be difficult to separate from time-evolving fluctuations.
Some argue that temperature variability is indeed increasing globally \cite{ref2,ref3}, whereas others conclude it is decreasing or remains practically unchanged \cite{ref4,ref40}. Meanwhile, a consensus appears to emerge that local variability in certain regions (e.g. Western Europe and North America) has indeed been increasing in the past 40 years \cite{ref5,ref6}.    
Here we investigate the nature of connections
between external forcing and climate variability conceptually by using a
laboratory-scale minimal model of mid-latitude atmospheric thermal
convection subject to continuously decreasing `equator-to-pole'
temperature contrast $\Delta T$, mimicking climate change. The analysis of temperature records from an
ensemble of experimental runs (`realisations') all driven by identical
time-dependent external forcing reveals that the collective
variability of the ensemble and that of individual realisations may be
markedly different -- a property to be considered when interpreting
climate records.
\end{abstract}

\flushbottom
\maketitle
%
%
\thispagestyle{empty}

\section*{Introduction}

To quantify connections between climate change and the temporal variability of
a climate index the typical procedure researchers follow is comparing
its recently observed fluctuations to those from a
base period \cite{baseperiod1,baseperiod2}. This approach is inherently built on the na\"ive assumption of ergodicity, a property
that does not apply to far-from-equilibrium processes. In
`climate-like' nonlinear, evolving systems the only way to
acquire appropriate expectation values -- as ``climate is what you
expect, weather is what you get''\cite{heinlein}-- would be ensemble averaging
over a multitude of parallel realisations of the system's response to
the same time-dependent forcing, all obeying the same physical laws
and differing only in their initial conditions. 
It is to be emphasized that differences between the ensemble members represent an inherent property of the problem, internal variability, and cannot only be associated with `measurement errors'. The
ensemble average of the paths of such parallel realisations in the
space of essential variables would then trace out a time-evolving,
so-called snapshot- or pullback- chaotic attractor \cite{ott, ghil1}.
It seems quite appropriate to adapt this approach to the description of any highly nonlinear chaos-like process, like e.g. turbulence. 

The concept's applicability in climatology has been demonstrated
in numerical models ranging from minimal models \cite{ghil1, ghil2, bodai} to
intermediate complexity GCMs \cite{plasim}, concluding that the snapshot
attractor framework provides the only self-consistent definition of
`climate' from the dynamical systems point of view. Obviously, for the
actual Earth system only a single
observable realisation exists but experiments in a laboratory
characterised by `climate-like' externally forced
dynamics can be repeated multiple times and thus provide a real world
test-bed for this approach, whose evaluation has so far been limited to numerical investigations.

\begin{figure}[t!]
\noindent\includegraphics[resolution=300,width=\textwidth]{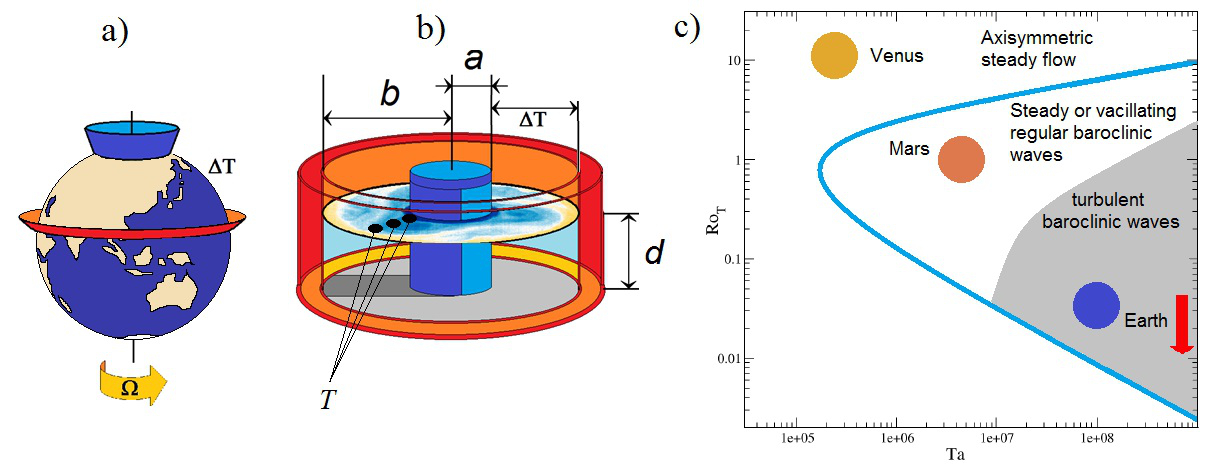}
\caption{Thermal convection in planetary atmospheres and in the laboratory. (a) Schematic diagram of the mid-latitude atmosphere of Earth, illustrating the basic boundary condition with a meridional temperature contrast $\Delta T$ between the warm equator (red) and a polar region (blue). The system is rotating at angular velocity $\Omega$. (b) Sketch of the differentially heated rotating annulus with its geometric parameters ($a=4.5$ cm, $b=12$ cm, $d=4.5$ cm) for which the boundary conditions are similar to those of the real atmosphere: warm outer rim (red), cold inner rim (blue). The locations of the three co-rotating thermometers, which were submerged by 0.5 cm into the bulk from above the water surface are also shown (black dots). The average of these signals at each time $t$ yielded `meridional mean temperature' $T (t)$. (c) Schematic regime diagram for rotating laterally heated systems\cite{read1, read2} in terms of thermal Rossby number $Ro_T \sim \Delta T/\Omega^2$ and Taylor number $Ta \sim \Omega^2/\nu^2$, where $\nu$ denotes the kinematic viscosity of the medium (for the precise formulation of these nondimensional parameters see Methods). The main flow regimes are indicated and the approximate positions of three planetary mid-latitude circulations are also shown\cite{read1}: $Ro_T^{\rm Venus} \gg 1$; $Ro_T^{\rm Mars}\approx 0.2$; $Ro_T^{\rm Earth}\approx 0.06$. The vertical arrow represents the trajectory of the dynamics in our experiment during the imposed `climate change' scenario: $Ta \approx 9.18\times 10^8$ stays constant, whereas thermal Rossby number decreases from $Ro_T \approx 0.041$ to $Ro_T \approx  0.013$.}
\label{fig1}
\end{figure}

The tabletop-size rotating, differentially heated annular wave tank we
use for this purpose is a widely studied experimental minimal model of
the mid-latitude Earth system\cite{read1, read2, bartos, benchmark} (Fig. \ref{fig1}a, Methods). It captures the
two essential components of large-scale atmospheric circulation:
lateral (`meridional') temperature difference and rotation. The
working fluid (de-ionised water) is located in the annular cavity
between two vertically aligned co-axial cylindrical side walls: the
one at the center (simulating the North Pole) is cooled, whereas the
rim (representing the equator) is heated with computer-controlled
thermostats. The tank is mounted on a turntable and
rotates around its axis of symmetry. The adjustable parameters (fluid
depth, rotation rate, temperature contrast) are set to yield
approximate dynamical similarity to the terrestrial atmosphere in
terms of thermal Rossby number, $Ro_T$, and Taylor number, $Ta$ (Fig. \ref{fig1}c, Methods)
\cite{read1, read2}. We log simultaneously (sampling rate 1 Hz, differential
resolution 0.05K) point-wise local temperature values via five digital
co-rotating thermometers, three of which penetrates into the free top
surface of the working fluid cavity from above, spaced uniformly along
a radius (Fig. \ref{fig1}b). The spatial average, $T_i(t)$, of these signals from three different `latitudes' is
used here as a surrogate for `meridional' mean temperature (index $i$ refers to the $i$-th ensemble member, i.e. experimental run).
{ Since there is no azimuthal bias in the annulus –- as there is, e.g. in the terrestrial atmosphere, due to land-ocean differences, topographical effects, etc. -- we would expect the statistical properties of temperature fluctuations to be the same at different azimuths. Thus, it is safe to assume that such a longitudinal average can also be considered a proper surrogate of the global average.} The other two identical sensors measure
the forcing temperatures at the center (inner cylinder) and in the
outer sidewall, whose difference $\Delta T$ quantifies the temperature
contrast driving the sideways convection.

The novelty of our experiments lies in the procedure of intentionally changing the
thermal boundary conditions in time, while keeping the rotation rate
fixed (so that a `day' i.e. one revolution of the tank takes $P = 3$
s). After a `base period' of ca. 2600 revolutions of constant $\Delta T$
the cooling element at the center is turned off. Following this abrupt
change in heat flux $T_i(t)$ is kept logged for another 3000 revolutions of time,
corresponding to a `global warming' scenario with gradually increasing
polar temperatures. { It is generally accepted that the North-South temperature contrast has been decreasing (and will continue to decrease) in the Northern Hemisphere due to climate change as reported, e.g. in the latest assessment report of IPCC\cite{IPCC}. The recent alarming findings\cite{EOS} about the rapidly melting Arctic also underline the existence of this phenomenon, showing twice as fast warming of the Arctic as that of the global mean.}

\begin{figure}[h!]
\noindent\includegraphics[resolution=300,width=\textwidth]{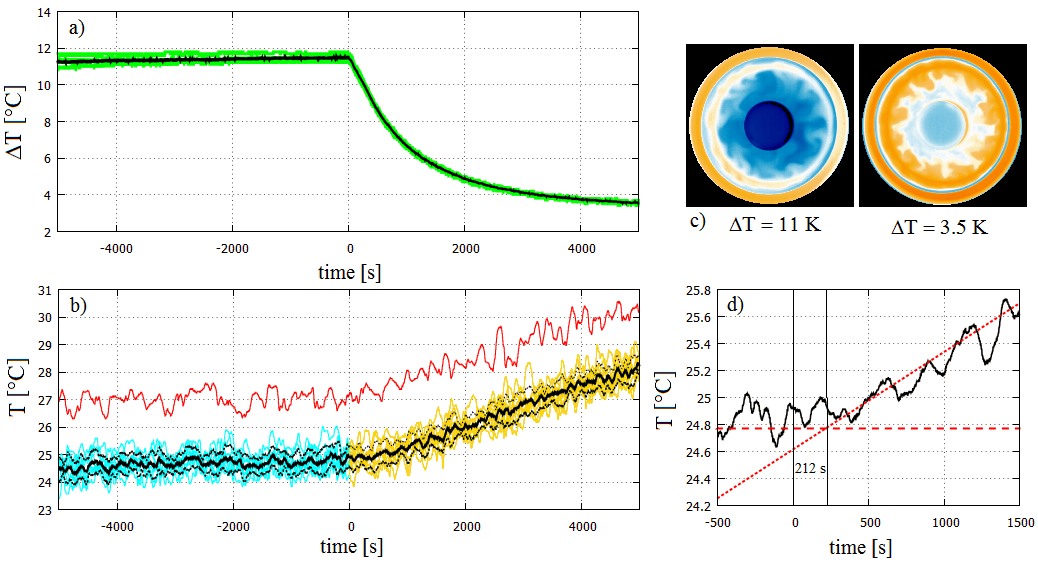}
\caption{Temperature trends and fluctuations in the experiment ensemble. (a) Temperature difference $\Delta T$ as measured between the outer and inner cylindrical sidewalls of the annular tank in all runs are shown (green) alongside their ensemble average at each time $t$ (black). (b) Time series $T (t)$ for each experimental realisation, coloured turquoise in the base period and orange in the `climate change' ($t> 0$) phase. One exemplary realisation of $T (t)$ is repeated in red and shifted above by $+2.5^\circ$C for better visibility. The ensemble average $ \langle T \rangle (t)$ of all nine realisations (black solid curve) and the corresponding $\pm 1 \sigma_e$ range (dotted black lines) is also indicated. (c) Two infrared thermographic snapshots (heat maps) of the surface temperature patterns (obtained in an additional control experiment) during the base period with temperature contrast $\Delta T = 11$ K (left) and toward the end of the `climate change' period (at $t = 4640$ s). Orange (blue) areas are warmer (colder) than average. (d) A blow-up of the ensemble average time series $ \langle T \rangle (t)$ showing linear trend lines fitted to the $t<0$ (dashed) and $t>0$ (dotted) periods. Their crossing point is found at $t = 212$ s that serves as an empirical measure of the delay time of the system's response {(baroclinic adjustment)} to the abrupt change in the forcing at $t = 0$.}
\label{fig2}
\end{figure}

\section*{Results}
Based on our criteria for the external forcing
sequence $\Delta T_i(t)$ to be accepted as `identical' (Methods) the analysis
was restricted to nine experimental runs and 10
000 s of continuous data from each of them with the onset of `climate change' (hereafter marked
as time zero, $t = 0$) occurring exactly at half time in all cases. The
forcing $\Delta T (t)$ in each considered realisation (Fig. \ref{fig2}a) follows an
exponential decay with characteristic timescale $\tau = 1085$ s for $t > 0$. The system's response $T_i(t)$ in each run, and even
their ensemble average $\langle T\rangle(t)$ shows significant fluctuations (Fig. \ref{fig2}b
and c) due to the geostrophic turbulent flow dominated by irregular
cyclonic (warm) and anticyclonic (cold) vortices \cite{bartos}.

\begin{figure}[h!]
\begin{center}
\noindent\includegraphics[resolution=300,width=0.66\textwidth]{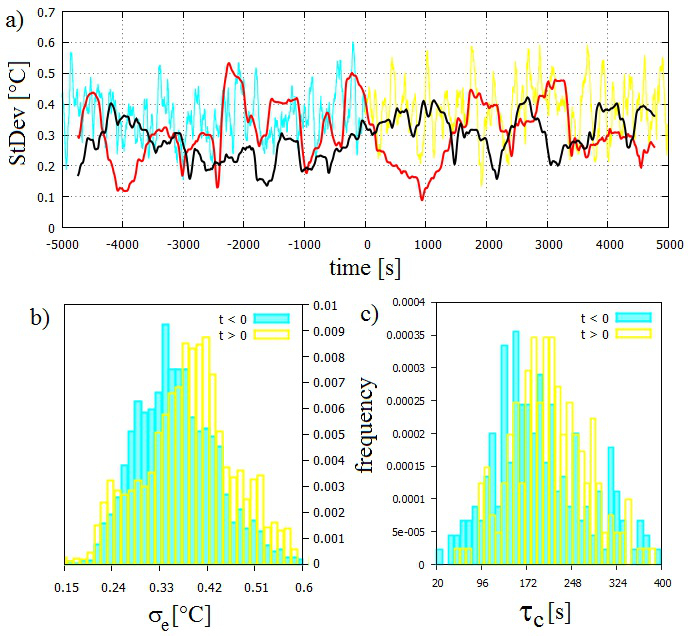}
\end{center}
\caption{Variances and correlation times of the `global mean temperature' time series. (a) The collective standard deviation $\sigma_e$ of the ensemble at each time $t$ (turquoise and orange curves) in the two periods (the colour coding is as in Fig. \ref{fig2}b). For comparison, time series of the 501-point running standard deviations of two experimental realisations are also shown (black and red curves) calculated after detrending with 4th order polynomials in both cases. (b) Histograms of the ensemble standard deviation $\sigma_e$ for the base period ($t<0$, turquoise) and the `climate change' period ($t>0$, orange). { (c) Histograms of the peak-to-peak time differences $\tau_c$ of the fluctuations pf $T_i(t)$ in the two periods, determined after 61-point running averaging and 5th order polynomial detrending. Here data from all individual realisations $T_i$ are combined. The colour coding is as in panels (a) and (b).}}
\label{fig3}

\end{figure}

Addressing variability in the system we first demonstrate the
difference between the `traditional' measures -- based on single
realisations -- and the ensemble statistics through the example of
standard deviations. We find that the centered running variances (within
501 s long windows) of the residuals of $T_i(t)$ following a
4-degree polynomial detrending in the different realisations may exhibit seemingly opposite
tendencies (Fig. \ref{fig3}a), and are thus not representative.
In the two chosen paths, one reaches the largest variability in the $t
< 0$ base period. Although the statistical comparison of the $t
< 0$ and $t > 0$ intervals yields no significant difference, the mean and
median indeed are somewhat smaller in the latter case (not shown). Thus -- if only
this particular record was known -- one could speculate that the
fluctuations of temperature generally decreased in the `climate change' phase
compared to the base period. The other exemplary case shows just the
opposite trend: a slight, statistically insignificant increase in mean
variability after $t = 0$. 

Meanwhile, in terms of the ensemble variance
$\sigma_e$, i.e. the standard deviation of the nine considered
realisations $T_i(t)\,\, (i=1,\dots,9)$ around $ \langle T \rangle (t)$ at each time instant $t$ (Fig.
\ref{fig3}a), the system's real sensitivity to changing $\Delta T$ is revealed (Fig.
\ref{fig3}b). The mean of $\sigma_e(t)$ shifts significantly by ca. 6.5\% from 0.35 to 0.38$^\circ$C at $t > 0$
and, more strikingly, the histogram changes from left-modal (skewness:
0.37) to right-modal (skewness: -0.10) after the initiation of
`climate change'. This result indicates that the paths of the realisations
differ from each other more in the presence of nonstationary forcing
than in the base period: even if the transition is hardly noticable in
the variance patterns of one single realisation its effect on the
whole ensemble is apparent.

{ The typical time difference $\tau_c$ between successive local extrema of the fluctuating temperature records $T_i(t)$ serves as a measure of the temporal variability of the `weather' in the system. The local maxima (minima) indicate the crossing of cyclonic (anticyclonic) eddies at the thermometer locations. We calculate the peak-to-peak time differences for each ensemble members, after a removal of a 5th order polynomial trend and applying a 61-point running mean for smoothing. The statistics of the obtained values of $\tau_c$ combined from all
experimental runs shows a significant shift when comparing the $t < 0$
and $t > 0$ periods; the mean increased from  199.2 s to  214.7 s (by around 8\%), the median
from  183 s to 209 s (by around 14\%).
This finding is consistent with the theoretical expectations: smaller $\Delta T$ yields
cyclonic and anticyclonic eddies of smaller size, scaling with the
so-called Rossby deformation radius $L_R$, i.e. proportionally to the square root of the imposed temperature gradient\cite{vallis}: $L_R \propto \sqrt{\Delta T}$. Whereas the drift velocity $c$ of baroclinic eddies is determined by the thermal wind balance and scales as $c\propto \Delta T^\mu$, where the exponent $\mu$ has been found to be between 0.88 and 1.17 in earlier experiments\cite{Fein,benchmark}, $\mu = 1$ being the theoretical value. Thus, the crossing timescale is expected to follow a $\tau_c \approx L_R/c \propto \Delta T^{0.5-\mu}$ dependence, yielding an increasing trend with decreasing $\Delta T(t)$ in time. Even in this respect, single-realisation statistics could be misleading:
due to the (geostrophic) turbulent nature of the flow the values of $\tau_c$ exhibit large variance in all cases that can easily suppress the slight trend.}

\begin{figure}[t!]
\noindent\includegraphics[resolution=300,width=\textwidth]{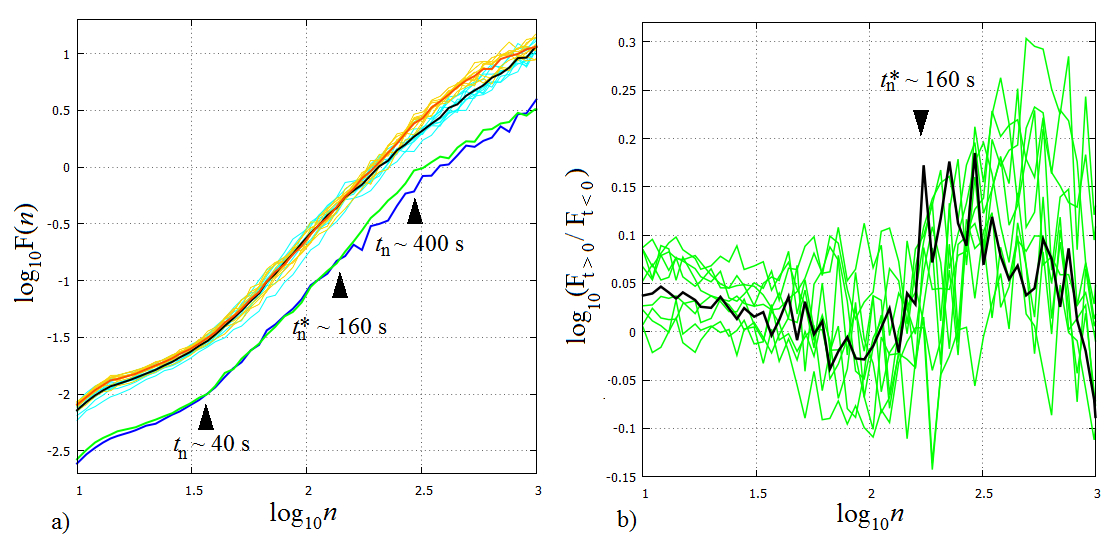}
\caption{Detrended fluctuation analysis of the `global mean temperature' ensemble. (a) The logarithm of DFA4 fluctuations $\log_{10}[F(n)]$ as a logarithm of window size $\log_{10}(n)$ ($n$ is the length of the window length in seconds). The spectra from the base period (turquoise curves) and from the `climate change' period (orange) exhibit indistinguishable behaviour (involving a scale break around $\log_{10}(n)=1.6$, corresponding to time scale $t_n \approx 40$ s) up to $\log_{10}(n)=2.2$ (i.e. $t^*_n \approx 160$ s), where the two sets of graphs detach and those corresponding to `climate change' reach larger fluctuations. The averages of the spectra are also plotted for the base period (black) and the `changing' period (red).
The DFA4 spectra of the ensemble average $ \langle T \rangle (t)$ in the base period (blue) and the `changing' period (green) show practically the same behaviour as the corresponding spectral averages. Upward shifting of these ensemble average spectra by $\log_{10}(3)\approx 0.477$ on the log-scale graph would yield practically identical curves to the aforementioned averages. (b) Amplification factors of the DFA4 fluctuations of the `climate change' period with respect to the base period in each individual run (green curves) and in the ensemble average $ \langle T \rangle (t)$ (thick black curve).}
\label{fig4}
\end{figure}

Further exploring temporal correlations we apply the method of detrended fluctuation analysis
(DFA)\cite{dfa,dfanature}, a strandard procedure for measuring the variability of a
signal around its local trend in time windows of length $n$ samples as a
function of $n$. DFA4 removes local (cubic) trends, thus it is more
suitable for our present purpose than e.g. Fourier transforms, since
DFA4 can readily handle nonstationarities (Methods). The DFA4 spectra
from all realisations (Fig. \ref{fig4}a) follow the same scaling properties,
exhibiting power law-type scaling with two scale breaks. Below $t_n
\approx 40$ s and above $t_n \approx 400$ s the scaling exponents are $\delta =
0.87$ and 1.1, respectively, implying $1/f$ noise-like correlated
fluctuations. Between these crossover points $\delta = 2.1$ is
found, characteristic for { geostrophic turbulence\cite{annulus_turb,japanok}: 
it can be shown that if the DFA4 spectra exhibit power-law scaling then the Fourier power spectrum of the time series in the frequency domain, $S(\omega)$ also does, following $S(\omega) \propto \omega^{-\beta}$, where $\beta=2\alpha - 1$ connects the two exponents\cite{dfa2}, yielding in the present case, $\beta \approx 3$. This is in good agreement with the theoretical result for isotropic geostrophic (two dimensional) turbulence\cite{2d_turb}. It is to be noted that the $\Delta T$-dependence of the exponent $\beta$ has been analyzed via comparing the ensemble-averaged power spectra of different (overlapping) sections of the time series $T_i(t)$, but no trend could be established (for more details, we refer to the Supplementary information). Thus, it can be stated that the `quality' of geostrophic turbulence did not change throughout the `climate change' period. 
}

Concerning
the differences between the stationary ($t < 0$) and `changing' ($t > 0$)
records (turquoise and orange curves in Fig. \ref{fig4}a, respectively), their
fluctuations up to a window size $t_n^* \approx 160$ s are perfectly
identical in the statistical sense. This is also apparent from the
averages of the two sets of spectra in Fig. \ref{fig4}a (red and black thick
curves). On the $t_n > t_n^*$ scale, however, the fluctuations of the
`changing' records are significantly larger. Note, that this timescale is still about an order of magnitude below $\tau=1085$ s i.e. the characteristic time of the `climate change' $\Delta T(t>0)$, but is of the same order as the empirical delay time of $\sim 200$ s of the dynamics estimated from the crossover point of the linear temperature trends of $ \langle T \rangle (t)$ in the two periods (Fig. \ref{fig2}d).  

Also shown are the DFA4 spectra of the ensemble averages $ \langle T \rangle (t<0)$ (blue line) and $ \langle T \rangle (t>0)$ (green line) following the same scaling and the same separation of the stationary and `changing' branches at $t_n^*$, as discussed above. Multiplying the fluctuation spectra of the ensemble averages by $\sqrt{N}=3$ ($N=9$ being the sample size of the ensemble) yields perfect match with the average of the single-realisation spectra. This property shows that the fluctuations of different realisations are perfectly uncorrelated on all time scales $t_n < \tau$: uncorrelated fluctuations average out following $1/\sqrt{N}$, whereas `ensemble-correlated' fluctuations would remain unaffected by the ensemble averaging. Here the latter are absent; no `collective variability' can be identified in the ensemble, despite of the identical forcing sequence $\Delta T(t)$. Obviously, on the time scale of $\tau$ collective behaviour does exist -- the trend itself -- but such large time windows are not sampled properly and are not evaluated in the spectra. The lack of collective fluctuations on the sub-$\tau$ scales highlights the largely nonlinear nature of the system's response to changing forcing. 

\begin{figure}[h!]
\begin{center}
\noindent\includegraphics[resolution=300,width=0.66\textwidth]{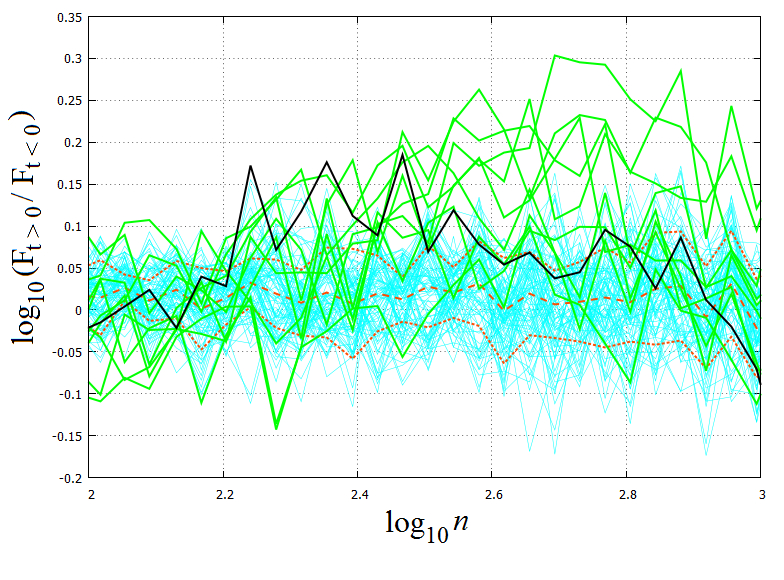}
\end{center}
\caption{Significance testing of the amplification factors of the DFA4 fluctuations. The amplification of the DFA4 fluctuations of the ensemble members and of the ensemble average are repeated from Fig. \ref{fig4}b with the same color coding in the $100\leq n \leq 1000$ domain. The turquoise curves indicate the 90 surrogate model time series. The average of the model spectra (red dashed curve) and the upper and lower bounds of the $\pm 3 \sigma$ intervals (red dotted curves) are also plotted.}
\label{fig5}
\end{figure}

The time-scale--dependence of variability amplification caused by `climate change' is visualized in Fig. \ref{fig4}b, where the ratios of DFA4 fluctuation spectra in the `changing' phase relative to the `base period' of the same run -- and those of the ensemble average -- are plotted. (Due to the logarithmic vertical axis this practically reflects the differences of the respective graphs in Fig. \ref{fig4}a.) Here again it becomes manifest that `climate change' does not affect the variability on the time scales below $t_n^*$ from the ensemble average point of view (the average amplification is close to zero), still, one can also easily spot individual realisations with either markedly increased or decreased variability in this spectral band as well. { Above $t_n^*$ all realisations exhibit clearly amplified variability. For the ensemble average it reaches a maximum increase of 47\% at around $t_n^*$ and stays around 20\% for the $t_n > t_n^*$ timescales up to $t_n \approx 800$ s.

To determine the statistical significance of the above results, we have carried out Monte Carlo statistical testing using a standard inverse-Fourier surrogate data method\cite{schreiber} (Methods). The null-hypothesis of the testing is that there are no fundamental changes in the dynamics of fluctuations between the $t<0$ `base period' and the $t>0$ `warming phase'. If this was the case, the fluctuations during the warming would exhibit very similar distribution and spectral properties as in the base period, superimposed onto a warming trend. In order to model this hypothesis, 10 model `warming' time series were created for each of the 9 ensemble members (i.e. 90 time series in total) using the Fourier amplitude spectra of their corresponding `base periods' but shuffling their phases. The resulting model series were then superimposed onto a polynomial warming trend, imitating the temporal development of the ensemble average $\langle T \rangle (t>0)$, shown in Fig. \ref{fig2}a, to yield a realistic increasing trend (Methods).

Comparing the DFA4 spectra of the model series to their respective base periods in the considered timescale range yields the `amplification factors' shown with turquoise curves in Fig. \ref{fig5}. The green curves corresponding to the actual ensemble members and the thick black curve denoting the ensemble average are repeated from Fig. \ref{fig4}b. The red dashed curve shows the mean of the model results and the dotted curves represent the $\pm 3\sigma$ interval. The vertical domain covered by the turquoise curves can be understood as a measure of variability that is due to finite-size effects of the imposed trend itself. It is apparent, however, that the measured ensemble data follow a markedly different distribution, thus the null-hypothesis in the considered timescale-range of $t_n > t_n^*$ can be rejected with a high confidence.  Towards the larger timescales, comparable to the typical eddy-crossing times $\tau_c$ and also to the characteristic time of baroclinic adjustment (as mentioned earlier), a clear increase of fluctuations can be observed, indicating real dynamical differences, not merely statistical artifacts.}

\section*{Discussion}

{ The present work provides, to the best of our knowledge, the first results from any laboratory experiment aiming to model the effects of climate change on mid-latitude atmospheric circulation. The authors do not claim that the lessons learned from the presented experimental minimal model could be directly applied or compared to the processes of the Earth system and the ongoing climate change. Perfect hydrodynamic similarity is impossible to achieve, thus the ratios between all of the relevant timescales (corresponding to the rotation, baroclinic adjustment, crossing time of cyclones, the changing of the temperature contrast $\Delta T$) cannot be set to scale properly. Nevertheless, the studied model as a dynamical system does share some important features with the climate system on the conceptual level: both are rotating, turbulent hydrodynamic systems, driven by the incoming differential heat fluxes, a forcing that changes in time. Due to the time-dependence of the forcing, these systems cannot reach an equilibrium state. Therefore, if one intends to survey the variability between the possible outcomes of such a process at any time instant, it is essential to consider a whole ensemble of realisations, subject to the same forcing scenario and differing only in their initial conditions.         

Despite of the large variability of the ensemble that is due to the nonlinear nature of the processes and the finite length of the studied records, the fluid dynamical interpretation of the observed flow phenomena is relatively straightforward. The system is in the state of well-developed geostrophic turbulence, that yields a power-law scaling in the power spectra of the fluctuations in both the wavenumber- and the frequency domain (Supplementary). The characteristic size of the cyclonic and anticyclonic eddies (corresponding to warm and cold temperature anomalies, respectively)  
tends to decrease as the `meridional' temperature gradient drops, in agreement with the theoretical expectations. In parallel, the zonal drift velocities decrease even faster during the process, therefore the characteristic timescale of `weather change' at a fixed measurement location increases significantly. This timescale is of the same order as the typical response time of the flow to the changes in the forcing (baroclinic adjustment) therefore fluctuations were found to increase markedly in this spectral band.  

``One experiment is no experiment'' has been the mantra of researchers for ages, but the idea behind the saying has always been the separation of measurement errors from significant signals. Here, however, the fluctuations are just as inherent, fully deterministic and dominant features of the underlying nonlinear processes -- just like in the Earth system -- as the large-scale trends themselves. The reason for the increasing ensemble variance lies in the system's extreme sensitivity to initial conditions -- a ubiquitous property of chaotic, long-range correlated systems. The authors firmly believe that the only proper approach for carrying out laboratory experiments on non-stationary turbulence would be conducting and systematically evaluating, ensembles of runs. In observational climatology this is not a viable option; we have only one Earth. Yet, the present experimental demonstration may help to increase awareness of the fact that a climate-like dynamical system can undergo a transition towards larger variability even without noticeable effects on the temporal fluctuations of one particular realisation. This message applies to the GCM community as well: climate variability information from a single numerical run (e.g. CO2 doubling scenario) could be misleading as it does not necessarily represent the full complexity of the underlying ensemble dynamics.      
}
\section*{Methods}         
\subsection*{Non-dimensional parameters, hydrodynamic similarity}
In large-scale environmental flows Rossby number $Ro\equiv U/(2|\Omega|L)$ -- with $U$ being the magnitude of the horizontal flow velocity, $L$ the horizontal extent of the domain and $\Omega$ the angular frequency of the planetary rotation -- quantifies the characteristic ratio of hydrodynamic acceleration and Coriolis acceleration. In the dynamics of atmospheric convection the thermal boundary conditions and the relationship $\rho(T)$ between the density and temperature of the fluid parcels are of fundamental importance just as well. A convenient nondimensional combination for quantifying all these factors is the thermal Rossby number Ro$_T$ (or Hide number), defined as
\begin{equation}
{\text {Ro}}_T=\frac{\alpha g d \Delta T}{\left(2\Omega\right)^2 L^2}\enspace,
\label{s2eq:12}
\end{equation}
where $\alpha$ is the 
coefficient of volumetric thermal expansion for the fluid, $d$ is the vertical scale, and $\Delta T$ is the `meridional' temperature contrast\cite{vallis}. For our calculations the annular gapwidth $b-a$ was taken as horizontal scale $L$ for the experiments.
Besides $Ro_T$ the kinematic viscosity $\nu$ of the medium also plays an important role in the dynamics; it introduces a `viscous cutoff' that dissipates too weak thermal winds and also damps the baroclinic instability of larger wavenumbers. This effect is parametrised by Taylor number $Ta$ that accounts for the ratio of rotational and viscous effects, and reads as
\begin{equation}
Ta=\frac{4\Omega^2 L^5}{\nu^2d}.
\label{Ta}
\end{equation} 
$Ro_T$ and $Ta$ are used in tandem to characterize the different dynamical regimes in rotating, thermally driven
systems, such as planetary atmospheres and their minimal models in the laboratory (Fig. \ref{fig1}c).

\subsection*{Experimental procedures, data selection}
For a detailed description of the experimental wave tank and the heating and cooling mechanism we refer to \cite{benchmark}. The temperature records were obtained using an ALMEMO temperature sensor array of NiCr sensors with a relative resolution of 0.05 K and 1 Hz sampling rate. The sensors were fixed onto a co-rotating mast above the free surface of the rotating annulus, and penetrated by 0.5 cm into the water surface. The data was transported in real-time via the co-rotating data aquisition module ALMEMO 8590-9, equipped with UHF/Bluetooth antenna. 
The initial temperature of the working fluid (de-ionised water) was set to $25\pm0.5^{\circ}$C before each measurement. After switching on the heating thermostats for the differential heating a transient period of 7600 s followed in order to reach quasi-equilibrium dynamics in each experimental run. Only after this period we started to log the data of the 5000 s long `base period'. The nine experimental runs considered in this work were selected based on the criterion that the forcing time series $\Delta T (t)$ of each realisation must not deviate by more than $0.3^\circ$C from the ensemble average $\langle\Delta T\rangle$ at any time $t$ (two of the original 11 experiments were thus excluded). The thermographic images of Fig. \ref{fig2}c were obtained by an InfraTec VarioCam infrared camera mounted above the set-up, operating in the spectral wavelength range of 7.5-14$\mu$m. These thermograms can be considered to represent surface temperature structures, since the penetration depth of this wavelength range into water is less than a millimeter. The images were taken during an additional experiment following the same forcing sequence, but with the thermometers removed from the working fluid for the sake of visibility. Therefore this run was not a member of the ensemble.      

\subsection*{Detrended fluctuation analysis}
DFA$p$\cite{dfa,dfa2} is a robust and easily implemented analysis of the temporal scaling properties of a fluctuating and non-stationary bounded time series $x_t$. Firstly a summation is applied to yield a cumulated (unbounded) time series $X_t$:
\begin{equation}
X_t = \sum_{i=1}^t[x_i-\langle x \rangle],
\end{equation}
where $\langle x \rangle$ denotes the mean of the time series. Next, the profile is divided into non-overlapping time windows $Y_j$ of length $n$ and for each a local least square polynomial fit $\xi(p-1)_j$ of order $p-1$ is calculated. Finally, the fluctuation is obtained as the root-mean-square deviation from the trend as
\begin{equation}
F(n)=\Big[\frac{1}{N}\sum_{j=1}^N(Y_j-\xi(p-1)_j)^2\Big],
\end{equation}
where $N$ is the number of $n$-sized windows of the time series. Note, that care must be taken to the fact that the congruence between $N$ and the length of the time series is often not zero. To preserve the remaining section the applied algorithm repeats the same dividing procedure from the end of $x_t$, thus, practically $2N$ segments are generated and the applied fluctuations are combined accordingly.
We determined the DFA$p$ fluctuation functions with $p = 2\dots 8$ for the time series $T_i(t)$ and observed that no significant differences appear between the spectra for $p>4$, therefore we limited our presentation of the results for the DFA4 computations only.

\subsection*{Surrogate data for the statistical testing}
{ The surrogate data for the model time series were generated using the
method developed by Schreiber and Schmitz and described in\cite{schreiber}.
The implementation of the algorithm is included in the open source  software package TiSeAn 3.0.1 for nonlinear time series analysis\cite{tisean} whose routine `surrogates' have been used for the present work. The principle of the method is the following: if the null hypothesis is true, the typical realizations of process are expected to share the
same power spectrum and amplitude distribution, thus such model time series need to be generated. This is carried out iteratively in the following procedure from the prescribed distribution and Fourier spectra of the actual data. First a sorted list of the values $\{x_n\}$ and the squared amplitudes of the Fourier transform of $\{x_n\}$, $S_k^2 = | \sum_{n=0}^{N-1}x_n \exp(i2\pi k n/N)|^2$ are obtained, where $N$ is the number of data points. 
Then a random shuffle of the data (without
replacement) $\{x_n^{(0)}\}$ is obtained. In a given iteration step, the shuffled data $\{x_n^{(i)}\}$ is brought to the desired sample power spectrum by taking the Fourier transform of $\{x_n^{(i)}\}$, replacing
the squared amplitudes $\{S_k^{2,(i)}\}$ by $\{S_k^2\}$ and then
transforming back. The phases of the complex Fourier
components are kept. This step enforces the correct
spectrum but usually the distribution will be modified.
Therefore, in the next step the resulting
series in rank–ordered to assume exactly the values taken
by $\{x_n \}$. Then, the spectrum of the resulting
$\{x_n^{(i+1)}\}$ will be modified again. These steps
have to be repeated several times;
at each iteration stage the remaining discrepancy
between the obtained and the desired spectra and distributions is checked and the iterations continue until a given accuracy
is reached. For finite $N$ a convergence
in the strict sense is not expected. Eventually, the transformation towards
the correct spectrum will result in a change which
is too small to cause a reordering of the values. Thus,
after rescaling, the sequence is not changed.

For each resulting model series, an increasing (5th order polynomial) warming trend was added. The properties of this warming trend were derived from fitting the polynomial formula to $\langle T \rangle (t)$ in the `climate change' period ($t > 0$). Thus, 90 model time series were obtained -- 10 for each ensemble member -- inheriting the power spectra and the rank-ordering of the original corresponding base period ($t < 0$) data.} 

\section*{Acknowledgements}
This research was funded by and conducted in the framework of the European High-performence Infrastructures in Turbulence (EuHIT) programme. U.H. and I.B. acknowledge financial support from the German Science Foundation (DFG) in the frame
of the Research Unit MS-GWaves (HA 2932/8-1)
The inspiring and fruitful discussions with Tam\'as T\'el, Imre M. J\'anosi and Anna Koh\'ari are highly acknowledged.

\end{document}